\documentclass{ws-mpla}
\usepackage[super]{cite}
\usepackage{graphicx}
\usepackage{hyperref}
\begin{document}

\markboth{Lixin Xu}
{Constraints on Dark Matter annihilation and Its Equation of State after {\it Planck} Data}

\catchline{}{}{}{}{}

\title{Constraints on Dark Matter annihilation and Its Equation of State after {\it Planck} Data}

\author{\footnotesize Lixin Xu\footnote{lxxu@dlut.edu.cn}}

\address{Institute of Theoretical Physics, School of Physics \&
Optoelectronic Technology, Dalian University of Technology, Dalian,
116024, P. R. China}

\maketitle

\pub{Received (Day Month Year)}{Revised (Day Month Year)}

\begin{abstract}
In this paper, the annihilation of dark matter $f_d\epsilon_0$ with nonzero equation of state $w_{dm}$ was studied by using the currently available cosmic observations which include the geometric and dynamic measurements. The constrained results show they are anti-correlated and are $w_{dm}=0.000390_{-0.000753}^{+0.000754}$ and $f_d\epsilon_0=1.172_{-1.172}^{+0.243}$ respectively in $1\sigma$ regions. With the including of possible annihilation of dark matter, no significant deviation from $\Lambda$CDM model was found in the $1\sigma$ region.

\end{abstract}

\ccode{PACS Nos.: 98.80.Cq, 95.35.+d, 98.70.Vc}

\section{Introduction}

The nature of dark matter (DM) is still one of the biggest puzzle in particle physics and cosmology. Its existence was confirmed by the observations from galaxy rotation curves, gravitational lensing, the large scale structure formation and cosmic microwave background (CMB) under the assumption that the Einstein's gravity theory is corrected. Actually, the cold dark matter plus a cosmological constant $\Lambda$, the so-called $\Lambda$CDM model, can almost agree with the most recent cosmic observations which include the type-Ia supernovae, the baryon acoustic oscillation (BAO) and CMB successfully at large scales. However, it has several potential problems on smaller scales \cite{Klypin1999,Moore1999,Zavala2009,Tikhonov2009,deBlok2001}. How to explain the discrepancies on large and small scale is currently still under debate \cite{Weinberg2013}. The warm dark matter has been proclaimed as a potential solution to the small scale difficulties of cold dark matter \cite{Zavala2009,Boylan-Kolchin2012,Lovell2012,Anderhalden2013,Menci2013,Papastergis2011}. Since the hot dark matter was ruled out due to the difficulty in forming the observed large scale structure, the remained focus point is whether the DM is cold or warm. If one takes the dark matter as a perfect fluid, its properties are characterized by its equation of state (EoS) $w_{dm}$ and effective sound speed $c^2_{s,eff}$ in its rest frame. And their values should be determined by the cosmic observations. A significant nonzero value of $w_{dm}$ indicates the dark matter is warm rather than cold. The sound speed determines the sound horizon of the fluid via the equation $l_s=c_{s,eff}/H$. The fluid can be smooth (or cluster) below (or above) the sound horizon $l_s$. If the sound speed is smaller, the perturbation of the fluid can be detectable on the relative large scale. In turn, the clustering fluid can influence the growth of density perturbations of matter, large scale structure and evolving gravitational potential which generates the integrated Sachs-Wolfe effects. Since the dark matter is responsible to forming the large scale structure of our Universe, we assume the effective speed of sound $c^2_{s,eff}=0$ in this work. Of course, one can extend to the case of nonzero $c^2_{s,eff}$ easily. Different values of $w_{dm}$ would change the background evolution history of our Universe. That means that the critical epoch, for example the equality time of matter and radiation, would be moved earlier or later; and the peaks of CMB temperature power spectrum would be increased or decreased; and the large scale structure formation history would be changed too. The comoving sound horizon $r_{s}$ at the recombination epoch was also changed, as a result the observations of BAO, the standard ruler, can be used to determine the values of $w_{dm}$. And the luminosity-redshift relation can also be used to constrain its values by treating Ia supernovae (SN) as standard candles due to the possible modification to the background evolution history trough the Hubble expansion rate. Then one can use the observed SN and BAO data sets to constrain the EoS of DM and fix the background evolution history from the geometric side. For the SN data points as "standard candles", the luminosity distances will be employed. In this paper, we keep to use the SNLS3 which consists of $472$ SN calibrated by SiFTO and SALT2, for the details please see \cite{ref:SNLS3}. Although the photometric calibration of the SNLS and the SDSS Supernova Surveys were improved \cite{ref:SNLS3recal}, they are still unavailable publicly. For the BAO data points as "standard ruler", we use the measured ratio of $D_V/r_s$, where $r_s$ is the comoving sound horizon scale at the recombination epoch, $D_V$ is the "volume distance" which is defined as $D_V(z)=[(1+z)^2D^2_A(z)cz/H(z)]^{1/3}$ where $D_A$ is the angular diameter distance. 

Due to the different properties of dark matter as indicated by different values of $w_{dm}$, here we have set the effective sound speed $c^2_{s,eff}$ to zero, the formation of the large scale structure and distribution of galaxies would be efficient indicators of this difference. Then the information from the large scale structure would be important to pin down the properties of dark matter from the dynamic side as a compensation to the geometric measurements such as SN and BAO. This is mainly because the fact that different cosmology models may have the same background evolution but the dynamic evolution would be different. That means that dynamic evolution is necessary to break the degeneracies between model parameters. Thanks to the measurements of WiggleZ Dark Energy Survey, a total $238,000$ galaxies in the redshift range $z<1$ were measured. These galaxies were split into four redshift bins with ranges $0.1<z<0.3$, $0.3<z<0.5$, $0.5<z<0.7$ and $0.7<z<0.9$. The corresponding power spectrum in the four redshift bins was measured; for details, please see \cite{ref:WiggleZ2012}. We estimate the nonlinear growth from a given linear growth theory power spectrum based on the principles of the halo model; actually the HALOFIT formula will be used in this paper \cite{ref:HALOFIT}. At the scale of halo, the growth of halos depends on the local physics, and not on the details of precollapse matter and the large scale distribution of matter. Thus, in the nonlinear regime the growth depends only on the nonlinear scale, the slope and curvature of the power spectrum \cite{ref:Bird2011}. In this work, we loosen the constraint to a zero equation of state and investigate the simplest model for dark matter, i.e. the one with a constant $w_{dm}$. Therefore, the main information is stored in the matter power spectrum. Also we assume the HALOFIT formula is still suitable for this case, though the formula would be modified due to the free-streaming of warm dark matter \cite{ref:Smith2011}. When the measurements from WiggleZ are used, the BAO data points from WiggleZ are not included. Because they come from the same galaxy sample as the $P(k)$ measurement. 

The initial conditions of the cosmological perturbations were fixed by the cosmic observations from CMB. Here the firs release of {\it Planck} data was employed due to the improvement of the quality of the cosmological data \cite{ref:Planck}. It allows us to give a tighter constraint to the cosmological parameter space when one uses the full information of CMB from the recently released {\it Planck} data sets which include the high-l TT likelihood ({\it CAMSpec}) up to a maximum multipole number of $l_{max}=2500$ from $l=50$, the low-l TT likelihood ({\it lowl}) up to $l=49$ and the low-l TE, EE, BB likelihood up to $l=32$ from WMAP9, the data sets are available on line \cite{ref:Planckdata}. 
  
On the particle physics side, the weakly interacting massive particle (WIMP) is a well motivated candidate. The direct detection experiments, such as DAMA \cite{ref:DAMA}, CoGeNT \cite{ref:CoGeNT} and CRESST \cite{ref:CRESST}, have suggested a low dark matter mass; for recent reviews, please see \cite{ref:WDMParticleReview}. The annihilation of the dark matter will release energy in the form of standard model particles. And the released energy would be absorbed by the surrounding gas, causing the gas to heat and ionize. Then the ionized fraction of electron $\chi_{ion}$ would be changed with respect to the redshift $z$. The excess free electrons interact with the CMB photons through the Thomson scattering, resulting a damping of the CMB power spectrum on small scales and a boost in the EE polarization power spectrum on large scales. Then through the measurements of CMB power spectrum, the properties of the dark matter can be detected indirectly. Several authors have studied the effects to the CMB power spectrum due to the decay and annihilation of the dark matter \cite{pierpaoli2004, chen_kamion2004,pad,zhang_etal2006,mapelli_etal2006,chuzhoy2008,arvi1,belikov_hooper2009,cirelli_etal2009,hutsi_etal2009,yuan_etal2009,arvi2,arvi3,slatyer_etal2009,kanzaki_etal2009,galli_etal2009,kim_naselsky2010,hutsi_etal2011,galli_etal2011,finkbeiner_etal2011,ref:ANatarajan,ref:Lopez2013,ref:Chluba2009}.  

In the literature, the equation of state of the dark matter was constrained from the different sides, for the recent results, please see Ref. \cite{ref:wdmxu} and references therein. However, we should notice that the evolution of the energy density of the dark matter with respect to the redshift $z$ is changed from $(1+z)^3$ to $(1+z)^{3(1+w_{dm})}$, when the equation of state of the dark matter $w_{dm}$ is not zero in the constant case. Then the equation of state is degenerated with annihilation of the dark matter particles through injected energy per unit volume at redshift $z$ \cite{ref:Chluba2009}
\begin{eqnarray}
\left.\frac{dE}{dVdt}\right|_{\chi\bar{\chi}}&=&2M_{\chi}c^2\langle\sigma v\rangle N_{\chi}N_{\bar{\chi}}\nonumber\\
&\approx&2.9\times 10^{-31}[1+z]^{6(1+w_{dm})}\text{eVs}^{-1}\text{cm}^{-3}\nonumber\\
&\times&\left[\frac{M_{\chi}c^2}{100\text{GeV}}\right]^{-1}\left[\frac{\Omega_{\chi}h^2}{0.13}\right]^{2}\left[\frac{\langle\sigma v\rangle}{3\times 10^{-26}\text{cm}^3/\text{s}}\right],
\end{eqnarray} 
which is defined as the energy liberated by the dark matter annihilations, where $M_{\chi}\equiv M_{\bar{\chi}}$ is the mass of the dark matter particle and its antiparticle; $\langle\sigma v\rangle$ is the thermally averaged product of the cross section and relative velocity of the annihilating the dark matter particles; and $N_{\chi}\equiv N_{\bar{\chi}}=N_{\chi,0}[1+z]^{3(1+w_{dm})}$ is the number density of the dark matter particles and their antiparticles, here $N_{\chi,0}\approx 1.4\times10^{-8}\text{cm}^{-3}\left[\frac{\Omega_{\chi}h^2}{0.13}\right]^{2}\left[\frac{2M_{\chi}c^2}{100\text{GeV}}\right]^{-1}$ is the present value of $N_{\chi}$. And this released energy will be deposited into the intergalactic medium (IGM), going into heating, and ionizations or excitations of atoms with a fraction $f_{d}(z)$ \cite{ref:Chluba2009}, i.e. 
\begin{equation}
\left.\frac{dE_d}{dVdt}\right|_{\chi\bar{\chi}}=f_{d}(z)\left.\frac{dE}{dVdt}\right|_{\chi\bar{\chi}}=f_{d}(z)\epsilon_0 N_H[1+z]^{3(1+w_{dm})}\text{eVs}^{-1}
\end{equation}
with the dimensionless parameter
\begin{equation}
\epsilon_0=1.5\times 10^{-24}\left[\frac{M_{\chi}c^2}{100\text{GeV}}\right]^{-1}\left[\frac{\Omega_{\chi}h^2}{0.13}\right]^{2}\left[\frac{\langle\sigma v\rangle}{3\times 10^{-26}\text{cm}^3/\text{s}}\right].
\end{equation}
Here $N_H\approx 1.9\times 10^{-7}\text{cm}^{-3}[1+z]^3$ is the number density of hydrogen nuclei in the Universe. In this work, we take $f_{d}$ as a constant in the range $[0,10]$. We modified the code {\it CosmoRec} \cite{ref:CosmoRec} to include a extra model parameter $w_{dm}$.

By a combination of CMB, SDSS BAO, SN and WiggleZ, the EoS of dark matter and $f_d\epsilon_0$  will be constrained. The plan of this paper is as follows: in section \ref{sec:BPEs}, we present the main background evolution and perturbation equations for dark matter with an arbitrary EoS. In section \ref{sec:results}, the constrained results are presented via the Markov Chain Monte Carlo (MCMC) method. Section \ref{sec:con} is the conclusion. 

\section{Background and Perturbation equations} \label{sec:BPEs}

For a nonzero value of the EoS of DM, the Friedmann equation for a spatially flat FRW universe reads
\begin{equation}
H^2=H^2_0\left[\Omega_{r}a^{-4}+\Omega_{b}a^{-3}+\Omega_{dm}a^{-3(1+w_{dm})}+\Omega_{\Lambda}\right],\label{eq:EHFE}
\end{equation}
where $\Omega_{i}=\rho_i/3 M^2_{pl}H^2$ are the present dimensionless energy densities for the radiation, the baryon, the dark matter and the cosmological constant respectively, where $\Omega_{dm}+\Omega_{r}+\Omega_{b}+\Omega_{\Lambda}=1$ is respected for a spatially flat universe.

In the synchronous gauge the perturbation equations of density contrast and velocity divergence for the dark matter are written as \cite{ref:Hu98,ref:wdmxu}
 \begin{eqnarray}
 \dot{\delta}_{dm}&=&-(1+w_{dm})(\theta_{dm}+\frac{\dot{h}}{2})+\frac{\dot{w}_{dm}}{1+w_{dm}}\delta_{dm}-3\mathcal{H}(c^2_{s,eff}-c^2_{s,ad})\left[\delta_{dm}+3\mathcal{H}(1+w_{dm})\frac{\theta_{dm}}{k^2}\right],\label{eq:wdmdelta}\\
\dot{\theta}_{dm}&=&-\mathcal{H}(1-3c^2_{s,eff})\theta_{dm}+\frac{c^2_{s,eff}}{1+w_{dm}}k^2\delta_{dm}-k^2\sigma_{dm}\label{eq:wdmv}.
 \end{eqnarray}
following the notations of Ma and Bertschinger \cite{ref:MB}, where $c^2_{s,ad}$ is the adiabatic sound speed of the dark matter
\begin{equation}
c^2_{s,ad}=\frac{\dot{p}_{dm}}{\dot{\rho}_{dm}}=w_{dm}-\frac{\dot{w}_{dm}}{3\mathcal{H}(1+w_{dm})}.
\end{equation}
In this work, we assume the shear perturbation $\sigma_{dm}=0$ and the adiabatic initial conditions. And for simplicity, we only consider a constant $w_{dm}$ in this work, although it is easy to be extended to the non-constant case. 

\section{Constrained Results} \label{sec:results}

To constrain the equation of state and annihilation of the dark matter from the currently available cosmic observations, on can use the Markov chain Monte Carlo (MCMC) method which is efficient in the case of more parameters. We modified the publicly available {\bf cosmoMC} package \cite{ref:MCMC} to include the perturbation evolutions of the dark matter with a general form of the equation of state according to the Eq. (\ref{eq:wdmdelta}) and Eq. (\ref{eq:wdmv}). The recombination was also modified to include the effect due to the possible nonzero equation of state of the dark matter. The  following eight-dimensional parameter space was adopted
\begin{equation}
P\equiv\{\omega_{b},\omega_c, \Theta_{S},\tau, f_d\epsilon_0,w_{dm}, n_{s},\log[10^{10}A_{s}]\}
\end{equation}
which priors are summarized in Table \ref{tab:results}. The new Hubble constant $H_{0}=72.0\pm3.0\text{kms}^{-1}\text{Mpc}^{-1}$ \cite{ref:hubble} was also adopted. The pivot scale of the initial scalar power spectrum $k_{s0}=0.05\text{Mpc}^{-1}$ is used in this paper.

\begin{center}
\begin{figure}[htb]
\includegraphics[width=10cm]{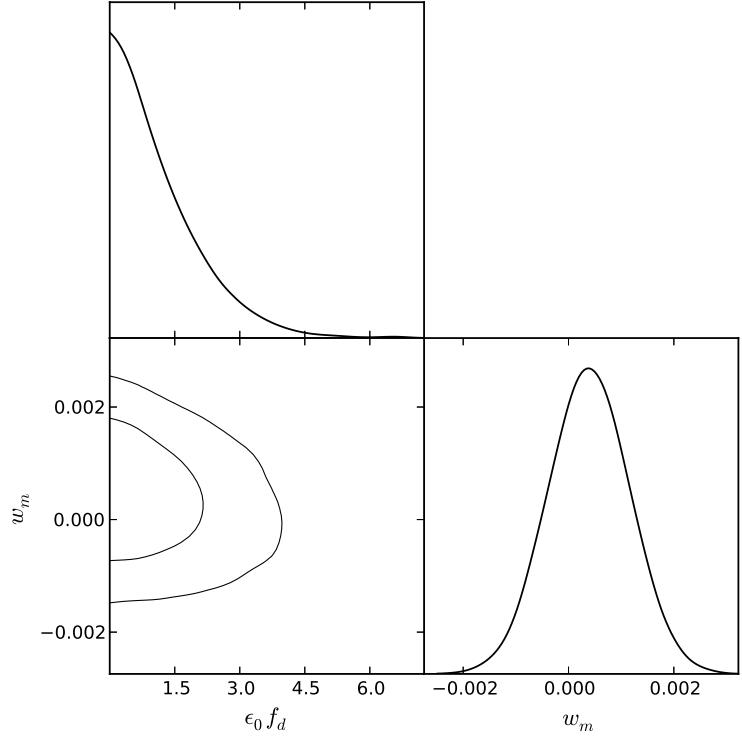}
\caption{The one-dimensional marginalized distribution on individual parameters and two-dimensional contours  with $68\%$ C.L., $95\%$ C.L. for $\Lambda$wDM model by using CMB+BAO+SN+WiggleZ data points.}\label{fig:wigglecontoure}
\end{figure}
\end{center}

Eight chains were run on the {\it Computing Cluster for Cosmos} for every cosmological models with different values of $w_{dm}$ and $f_d\epsilon_0$ with priors which are gathered in the second column of Table \ref{tab:results}. The chains are stopped when the Gelman \& Rubin $R-1$ parameter is $R-1 \sim 0.02$ which guarantees the accurate confidence limits. The constrained results are summarized in Table \ref{tab:results}. The result is compatible to our previous result obtained in Ref. \cite{ref:wdmxu}. Correspondingly, the one-dimensional and two-dimensional contours for $\epsilon_0 f_d$ and $w_{dm}$ was plotted in Fig. \ref{fig:wigglecontoure}. 
\begin{center}
\begin{table}[tbh]
\tbl{The mean values with $1,2,3\sigma$ errors and the best fit values of model parameters, where SNLS3, BAO, {\it Planck}+WMAP9 and WiggleZ measurements of matter power spectrum are used.}
{\begin{tabular}{@{}cccc@{}} \toprule
\hline\hline Prameters & Priors & Mean with errors & Best fit \\ \hline
$\Omega_b h^2$ & $[0.005,0.1]$ & $0.0222_{-0.000302-0.000575-0.000743}^{+0.000300+0.000602+0.000794}$ & $0.0224$\\
$\Omega_c h^2$ & $[0.01,0.99]$ & $0.117_{-0.00158-0.00311-0.00411}^{+0.00158+0.00317+0.00414}$ & $0.117$\\
$100\theta_{MC}$ & $[0.5,10]$ & $1.0414_{-0.000579-0.00117-0.00157}^{+0.000584+0.00112+0.00148}$ & $1.0414$\\
$\tau$ & $[0.01,0.8]$ & $0.0877_{-0.0138-0.0242-0.0311}^{+0.0124+0.0262+0.0355}$ & $0.0958$\\
$f_d\epsilon_0$ & $[0,10]$ & $1.172_{-1.172-1.172-1.172}^{+0.243+1.966+3.258}$ & $0.739$\\
$w_{m}$ & $[-0.2,0.2]$ & $0.000390_{-0.000753-0.00147-0.00196}^{+0.000754+0.00149+0.00197}$ & $-0.000300$\\
$n_s$ & $[0.5,1.5]$ & $0.972_{-0.0103-0.0177-0.0201}^{+0.00762+0.0189+0.0290}$ & $0.973$\\
${\rm{ln}}(10^{10} A_s)$ & $[2.4,4]$ & $3.122_{-0.0404-0.0685-0.0797}^{+0.0309+0.0731+0.106}$ & $3.125$\\
\hline
$\Omega_\Lambda$ & - & $0.705_{-0.0110-0.0230-0.0309}^{+0.0121+0.0216+0.0282}$ & $0.699$\\
$\Omega_m$ & - & $0.295_{-0.0121-0.0216-0.0282}^{+0.0110+0.0230+0.0309}$ & $0.301$\\
$\sigma_8$ & - & $0.847_{-0.0233-0.0451-0.0590}^{+0.0235+0.0469+0.0627}$ & $0.828$\\
$z_{re}$ & - & $10.787_{-1.0959-2.151-2.826}^{+1.0891+2.157+2.878}$ & $11.458$\\
$H_0$ & - & $68.880_{-0.972-1.919-2.528}^{+0.968+1.939+2.549}$ & $68.259$\\
$Y_P$ & - & $0.248_{-0.000129-0.000248-0.000323}^{+0.000129+0.000256+0.000336}$ & $0.248$\\
$10^9 A_s e^{-2\tau}$ & - & $1.906_{-0.0715-0.0941-0.108}^{+0.0340+0.116+0.170}$ & $1.879$\\
${\rm{Age}}/{\rm{Gyr}}$ & - & $13.750_{-0.0620-0.121-0.158}^{+0.0619+0.123+0.163}$ & $13.797$\\
\hline\hline
\end{tabular}\label{tab:results}}
\end{table}
\end{center}

\begin{center}
\begin{figure}[htb]
\includegraphics[width=10cm]{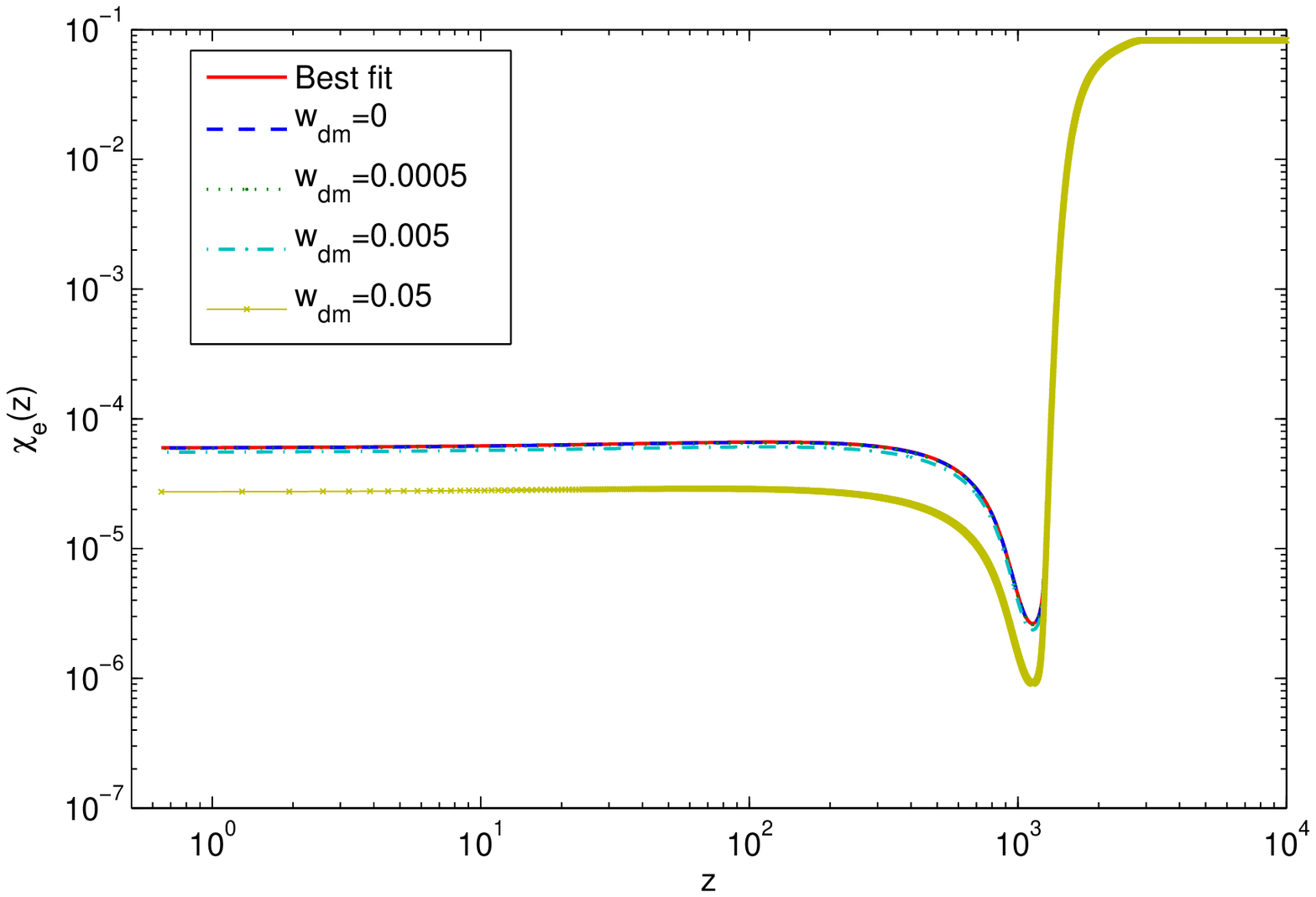}
\includegraphics[width=10cm]{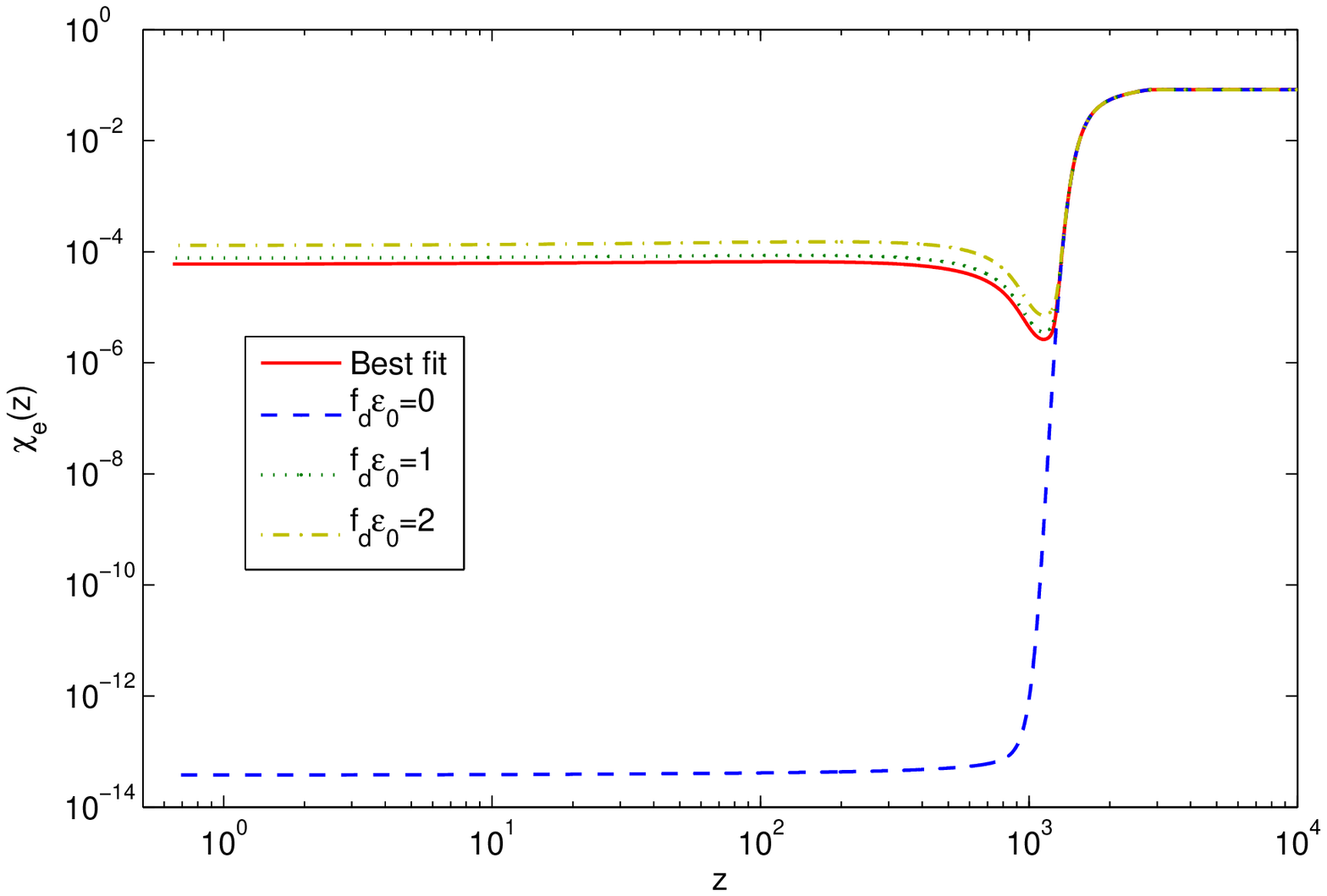}
\caption{The ionized fraction of electron with respect to the redshift for different values of $w_{dm}$, where the other relevant cosmological model paramours were fixed to their best fit values as listed in Table \ref{tab:results}.}\label{fig:chiez}
\end{figure}
\end{center}

To show the effects of $w_{dm}$ and $f_d\epsilon_0$ to the ionized fraction of electron $\chi_e(z)$, we fix the other relevant cosmological parameters to their best fit values as given in Table \ref{tab:results} and vary the values of $w_{dm}$ and  $f_d\epsilon_0$ around their best fit values. Their effects on the ionization due to different values of $w_{dm}$ and $f_d\epsilon_0$ are plotted in Figure \ref{fig:chiez}. As shown in this figure, large values of $w_{dm}$ will lower the ionized fraction of electron. And the annihilation of the dark matter will increase the ionized fraction of electrons as expected. Just due to the annihilation of the dark matter, the ionized fraction of electron is increased instead of downing to almost zero. Therefore, it can be easily understand that the comic observations of CMB data can be used to constrain the annihilation of the dark matter. 

\section{Conclusion} \label{sec:con}

In this paper, the degeneracy between the equation of state and annihilation of the dark matter was studied by using the currently available cosmic observations which include SNLS, SDSS and BAO to fix the background evolution and the CMB of the first 15.5 months data from {\it Planck} to fix the initial conditions of the perturbations and the WiggleZ measurement of power spectrum to fix the large structure formation. We have found the latest data sets provide that constraint $w_{dm}=0.000390_{-0.000753-0.00147-0.00196}^{+0.000754+0.00149+0.00197}$ and $f_d\epsilon_0=1.172_{-1.172-1.172-1.172}^{+0.243+1.966+3.258}$ in $3\sigma$ regions. The current data show also the anti-correlation between $w_{dm}$ and $f_d\epsilon_0$. With the including of possible annihilation of DM, no significant deviation from $\Lambda$CDM model is found in the $1\sigma$ region.  

\section*{Acknowledgments}
L. Xu thanks the invitation of Prof. M. Yu. Khlopov to give contribution to this special issue of MPLA. L. Xu's work is supported in part by NSFC under the Grants No. 11275035 and "the Fundamental Research Funds for the Central Universities" under the Grants No. DUT13LK01.

\end{document}